\documentclass[prd,reprint,superscriptaddress]{revtex4-1}

\usepackage{graphics}
\usepackage{amssymb,amsmath}
\usepackage{color}
\usepackage{wasysym}
\usepackage[tight]{subfigure}
\usepackage{booktabs}
\usepackage{multirow}
\usepackage{rotating}
\usepackage{float}
\usepackage{slashed}
\usepackage{ulem}
\usepackage{array}
\usepackage{slashed}

\newcommand{\beq}{\begin{equation}}
\newcommand{\eeq}{\end{equation}}
\newcommand{\beqa}{\begin{eqnarray}}
\newcommand{\eeqa}{\end{eqnarray}}

\newcommand{\hc}{\text{h.c.}}
\newcommand{\nn}{\nonumber}

\begin{document}
\title{Inverse Seesaw and Portal Dark Matter}
\author{Chakrit Pongkitivanichkul}
\email{chakpo@kku.ac.th}
\affiliation{Department of Physics, Khon Kaen University, 123 Mitraphap Rd., Khon Kaen, 40002, Thailand}
\affiliation{National Astronomical Research Institute of Thailand, Chiang Mai 50180, Thailand}
\author{Nakorn Thongyoi}
\email{nakorn$\_$t@kkumail.com}
\affiliation{Department of Physics, Khon Kaen University, 123 Mitraphap Rd., Khon Kaen, 40002, Thailand}
\author{Patipan Uttayarat}
\email{patipan@g.swu.ac.th}
\affiliation{Department of Physics, Srinakharinwirot University, 114 Sukhumvit 23rd Rd., Wattana, Bangkok, 10110, Thailand}
\affiliation{National Astronomical Research Institute of Thailand, Chiang Mai 50180, Thailand}

\begin{abstract}
We study the phenomenology of the inverse seesaw mechanism in the scalar-Higgs portal dark matter model. The model is an extension of the Standard Model including two additional neutrinos, a singlet scalar and a fermionic dark matter. We consider the inverse seesaw mechanism where the mass of 2 additional neutrinos are made dynamic by the singlet scalar. We found that the natural scale for the scalar vacuum expectation value is naturally close to the weak scale. Motivating by this fact, we focus on the possibility of the singlet scalar connecting with dark matter, i.e., the scalar is also the mediator between dark sector and the Standard Model. We perform a numerical analysis over the parameter space subject to the indirect and direct detection constraints. The feasible region of the parameter space will be discussed.
\end{abstract}

\maketitle

%%%%%%%%%%%%%%%%%%%%%%%%%%%%%%%%%%%%%%%%%%%%%%%%%%%%%%%%%%%%%%%%%%%%%%%%%%%%%%%%%%%%%%%%%%%%%%%%%%%%%%%%%%%%%%%%%%%
%%%%%%%%%%%%%%%%%%%%%%%%%%%%%%%%%%%%%%%%%%%%%%%%%%%%%%%%%%%%%%%%%%%%%%%%%%%%%%%%%%%%%%%%%%%%%%%%%%%%%%%%%%%%%%%%%%%
%%%%%%%%%%%%%%%%%%%%%%%%%%%%%%%%%%%%%%%%%%%%%%%%%%%%%%%%%%%%%%%%%%%%%%%%%%%%%%%%%%%%%%%%%%%%%%%%%%%%%%%%%%%%%%%%%%%
%%%%%%%%%%%%%%%%%%%%%%%%%%%%%%%%%%%%%%%%%%%%%%%%%%%%%%%%%%%%%%%%%%%%%%%%%%%%%%%%%%%%%%%%%%%%%%%%%%%%%%%%%%%%%%%%%%%
%%%%%%%%%%%%%%%%%%%%%%%%%%%%%%%%%%%%%%%%%%%%%%%%%%%%%%%%%%%%%%%%%%%%%%%%%%%%%%%%%%%%%%%%%%%%%%%%%%%%%%%%%%%%%%%%%%%
%%%%%%%%%%%%%%%%%%%%%%%%%%%%%%%%%%%%%%%%%%%%%%%%%%%%%%%%%%%%%%%%%%%%%%%%%%%%%%%%%%%%%%%%%%%%%%%%%%%%%%%%%%%%%%%%%%%

\section{Introduction}

The discovery of the Higgs boson in 2012, together with decades of electroweak precision tests, have been hailed as the remarkable success of the Standard Model (SM) of particle physics. However, the existence of neutrino masses and dark matter (DM) strongly suggests an extension beyond the Standard Model (BSM) which requires new degrees of freedom. The connection between these new physics is therefore simplistic yet tantalizing. %The link between dark matter and neutrino masses is an attractive possibility. 

Seesaw mechanisms are considered the best explanation for the smallness of the neutrino mass.  In the minimal realisation of the seesaw mechanism, a right handed neutrino is introduced to SM where the active left-handed neutrino gains its mass from its mixing with the right-handed neutrino. The mass can be obtained from the formula $m_{\nu} = \frac{m_D^2}{M}$ where $m_D$ is the Dirac mass and $M$ is the Majorana mass of the right-handed neutrino. In seesaw mechanism, the sub-eV neutrino mass requires $M \sim 10^{16}$ GeV. This huge difference between the electroweak scale and the seesaw scale leads to a strong suppression for any potential phenomenological signals from accelerator experiments and astronomical observations. Although the minimal seesaw mechanism provides an interesting explanation for neutrino mass, it certainly lacks the testability and thus diminishing the chance for connecting the origin of neutrino mass and DM observables.

To bring the seesaw scale closer to the electroweak scale, one can employ the so-called inverse seesaw mechanism. Inspired by String/M theory, the SM is extended by 2 sterile neutrinos and an electroweak scalar singlet~\cite{Mohapatra:1986bd}. It has been shown that the small neutrino mass can be generated from new physics around TeV scale. The connection with DM within the inverse seesaw context has recently gained interest~\cite{Huang:2014bva,PhysRevD.90.065034,Escudero:2016tzx,Escudero:2016ksa,DeRomeri:2017oxa}. It is well known that DM cannot take part directly in the seesaw mechanism~\cite{ISHIDA2014242,Lee:1977tib,Marciano:1977wx,Petcov:1976ff,Pal:1981rm}. If the sterile neutrino is the DM, it would decay into gamma rays and active neutrino. On the other hand, the option with the singlet scalar being DM is also limited due to its mixing with the Higgs which leads to the shorter lifetime. To avoid such pitfalls, one can instead utilise the heavy neutrinos or the singlet scalar field as a mediator to connect with dark sector~\cite{DeRomeri:2017oxa,Primulando:2017kxf}.

In this paper, we are interested in exploring the possibilities of connecting DM to the inverse seesaw model. In particular, we will consider the model in which the Dirac mass for the additional sterile neutrinos $M_D$ is explained by the dynamic of a scalar field mediator. The mediator is then connected to the fermionic dark sector via scalar and pseudo scalar coupling. The paper is organised as follows. First we provide the set up of the model in Sec.~\ref{ch:model}. In this section, the inverse seesaw mechanism where the lightest neutrino being identified with the SM active neutrino is described. The neutrino couplings are derived and the scalar mediator mixing with Higgs is investigated. Phenomenology and the constraints on the model is presented in Sec.~\ref{ch:pheno}. The invisible $Z$ boson decay is discussed. The indirect detection via gamma ray and neutrino telescope is investigated. Then the direct detection via nucleon scattering is studied. The scan of all parameter space of the model subject to constraints is shown in Sec.~\ref{ch:results}. We finally conclude in Sec.~\ref{ch:con}.

%%%%%%%%%%%%%%%%%%%%%%%%%%%%%%%%%%%%%%%%%%%%%%%%%%%%%%%%%%%%%%%%%%%%%%%%%%%%%%%%%%%%%%%%%%%%%%%%%%%%%%%%%%%%%%%%%%%
%%%%%%%%%%%%%%%%%%%%%%%%%%%%%%%%%%%%%%%%%%%%%%%%%%%%%%%%%%%%%%%%%%%%%%%%%%%%%%%%%%%%%%%%%%%%%%%%%%%%%%%%%%%%%%%%%%%
%%%%%%%%%%%%%%%%%%%%%%%%%%%%%%%%%%%%%%%%%%%%%%%%%%%%%%%%%%%%%%%%%%%%%%%%%%%%%%%%%%%%%%%%%%%%%%%%%%%%%%%%%%%%%%%%%%%

\section{The model} \label{ch:model}

In order to construct the model with the inverse seesaw mechanism, we extend the SM by adding 2 additional fermions: the right-handed $N_1$ and the left-handed $N_2$. We also add an electroweak singlet scalar $\Phi$ whose vacuum expectation value (vev) is responsible for the Dirac mass for the new fermions.
In addition to the SM gauge groups, the $Z_2$ discrete symmetry where all the SM fields are even is imposed, in order to obtain the neutrino mass matrix texture desired by the model, see Tab.~\ref{tab:contents}.
\begin{table}
\begin{tabular}{| c | c | c | c |}
\hline
& $SU(2)_L$ & $U(1)_Y$ & $Z_2$\\
\hline
$L$ & $\mathbf{2}$ & $-1/2$ & $+1$\\
$H$ & $\mathbf{2}$ & $1/2$ & $+1$\\
$N_1$ & $1$ & $0$ & $+1$\\
$N_2$ & $1$ & $0$ & $-1$\\
$\Phi$ & $1$ & $0$ & $-1$\\
$\chi$ & $1$ & $0$ & $-1$\\ \hline
\end{tabular}
\caption{The field contents and their transformation properties.}
\label{tab:contents}
\end{table}

In a 4-component notation, the Lagrangians for the neutrino sector and the scalar sector are 
\begin{align}
\mathcal{L}_{\rm{N}} =& -y\overline{L}\widetilde{H} N_{\text{1R}}-g\Phi \overline{N_{\text{2L}}}N_{\text{1R}}\nn\\
&\quad-\frac{\mu_N}{2}(\overline{N_{1R}^{c}}{N}_{1R}+\overline{N_{2L}^{c}}{N}_{2L})+\text{h.c.}\label{eq_neu}, \\
\mathcal{L}_{scalar}&=(D_{\mu}H)^{\dagger}(D^{\mu}H)+\frac12(\partial_{\mu}\Phi)(\partial^{\mu}\Phi)\nn\\
&\quad-V(H,\Phi),
\end{align}
where
\begin{align}
V(H,\Phi) =&-\mu^{2}H^{\dagger}H + \lambda(H^{\dagger}H)^{2}-\frac{\mu_{\phi}^{2}}{2}\Phi^2+\frac{\lambda_{\phi}}{4}\Phi^{4}\nn\\
&\quad+\frac{\lambda_{\phi H}}{2}\Phi^2 H^{\dagger}H \label{eq_neu2}.
\end{align}
In the above equations we have $\widetilde{H}=i\sigma_{2}H^{*}$ and $\psi^{c}=C\overline{\psi}^{T}$, $\mu_N$ is Majorana mass of the heavy neutrino. Note the $\mu_N$ term would violate the U(1) lepton number symmetry. Thus we expect it to be small. Moreover, the smallness of $\mu_N$ is technically natural by 't Hooft's naturalness principle. Lastly we can extend our model such that small $\mu_N$ is generated in similar fashion to models considered in ~\cite{Huang:2014bva,Ahriche:2016acx}. This model is quite different than the recent study of~\cite{DeRomeri:2017oxa} where the dynamic part of the inverse seesaw is in the Majorona term. 

The potential in Eq.~\eqref{eq_neu2} admits non-trivial vacuum expectation values (vev) for the two scalar fields. We can expand both $H$ and $\Phi$ around their vev as
\begin{eqnarray}
H = \frac{1}{\sqrt{2}}\begin{pmatrix} w^{+} \\ h' +iz+ v \end{pmatrix}, \quad\Phi = \phi' + v_{\phi}, %\Phi = \frac{\widetilde{\phi} + v_{\phi}}{\sqrt{2}},
\end{eqnarray}
where $w^+$ and $z$ are the would be Goldstone bosons eaten by the $W^+$ and $Z$ gauge bosons. Thus we see that the scalar sector contains two real degrees of freedom, denoted by $h'$ and $\phi'$.

Due to the scalar mixing with the Higgs, massive $N_1, N_2$ and $\phi'$ are not stable and hence cannot be a good DM candidate. We assume that DM resides in a separated sector which is connected to our sector by the scalar $\Phi$. We will further assume for simplicity that DM is a fermion.  Thus Lagrangian for DM is given by
\begin{equation}
	\mathcal{L}_{DM} = \Phi\overline{\chi}(G+i\tilde{G}\gamma^5)\chi + M\overline{\chi}\chi,
	\label{eq:darkmatter}
\end{equation}
where $G$ is a coupling and $\tilde{G}$ is a pseudo-scalar coupling. Note that this dark sector  explicitly breaks the $Z_2$ symmetry and contains the CP-violation which will be mediated to the neutrino sector.
%%%%%%%%%%%%%%%%%%%%%%%%%%%%%%%%%%%%%%%%%%%%%%%%%%%%%%%%%%%%%%%%%%%%%%%%%%%%%%%%%%%%%%%%%%%%%%%%%%%%%%%%%%%%%%%%%%%
%%%%%%%%%%%%%%%%%%%%%%%%%%%%%%%%%%%%%%%%%%%%%%%%%%%%%%%%%%%%%%%%%%%%%%%%%%%%%%%%%%%%%%%%%%%%%%%%%%%%%%%%%%%%%%%%%%%

%---------------------------------------------
\subsection{Neutrino mass}
\label{ch:neumass}

\begin{figure}
\centering
\includegraphics[width=0.95\linewidth]{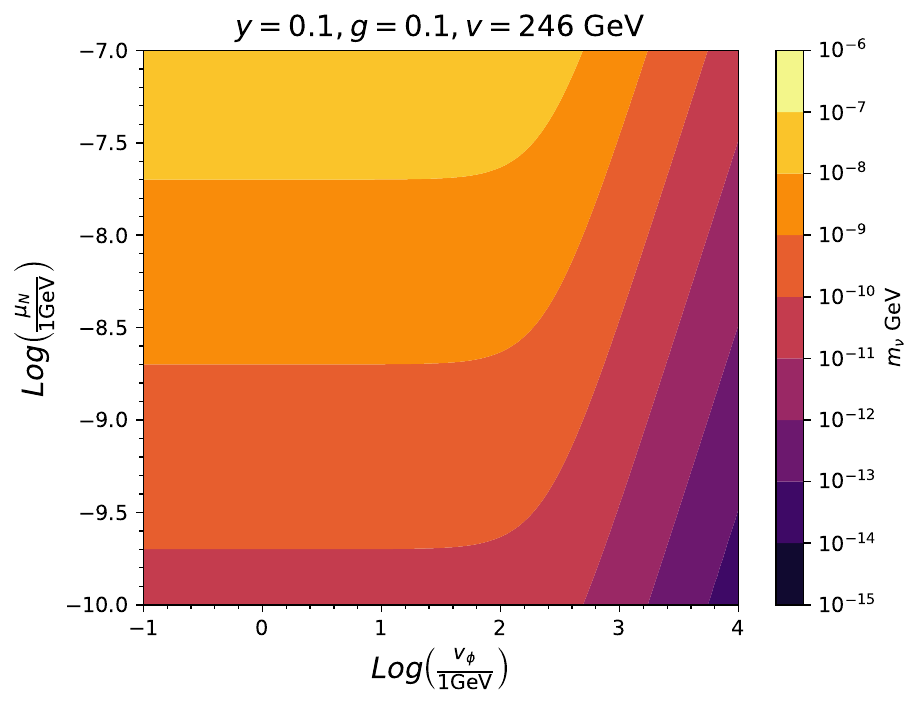}
\caption{The contour plot showing the value of $\log_{10}\left( \frac{m_{\nu}}{1 {\rm GeV}}\right)$.}
\label{fig:neucon}
\end{figure}

Neutrino masses arise from the Yukawa interactions in  Eq.~(\ref{eq_neu})%.  Expanding the fields around their vacuum expectation values %{\color{red}[notational consistency!]}
%\begin{eqnarray}
%H = \frac{1}{\sqrt{2}}\begin{pmatrix} w^{+} \\ h +iz+ v \end{pmatrix}, \quad\Phi = \phi + v_{\phi} %\Phi = \frac{\widetilde{\phi} + v_{\phi}}{\sqrt{2}}
%\end{eqnarray}
%one gets
\begin{eqnarray}
	\mathcal{L}_{\text{N}}&\supset& -\frac{1}{2\sqrt{2}}yv\left(\overline{\nu_{L}}N_{\text{1R}}+\overline{N_{\text{1R}}^{\text{c}}}\nu_{\text{L}}^{c}\right)\nonumber\\
	& &-\frac{1}{2}gv_{\phi}\left(\overline{N_{\text{2L}}}N_{\text{1R}}+\overline{N_{\text{1R}}^{c}}N_{\text{2L}}^{c}\right)\nonumber\\
	& & -\frac{\mu_N}{2}(\overline{N_{1R}^{c}}{N}_{1R}+\overline{N_{2L}^{c}}{N}_{2L})+\text{h.c.}
\end{eqnarray}
where we used the fact that 
$\overline{\nu_{\text{L}}}N_{\text{1R}}=\overline{N_{\text{1R}}^{c}}\nu_{\text{L}}^{c}$.
The Lagrangian for neutrino sector becomes the mass matrix under the basis $\psi_{R} = (\nu_{\text{L}}^{c}, N_{\text{1}}, N_{2}^{c})$
\begin{equation}
\mathcal{L}_{\text{N}}^{\text{Mass}} = - \frac{1}{2}\overline{\psi_{R}^{c}} M \psi_{R}
\end{equation}
where
\begin{align}
M = \begin{pmatrix} 0 & y v/\sqrt{2} & 0 \\ y v/\sqrt{2} & \mu_N & g v_{\phi} \\ 0 & g v_{\phi} & \mu_N \end{pmatrix} \label{eq:neumass}
\end{align}
We can diagonalize the mass matrix by an $SO(3)$ rotation matrix $R$
\begin{equation}
	\begin{pmatrix}\nu^c_l\\N_{1R}\\N_{2L}^c\end{pmatrix} %= R \begin{pmatrix}\nu_l^{c\prime}\\N_{1R}^{\prime}\\N_{2L}^{c\prime}\end{pmatrix} 
		\equiv R\begin{pmatrix}\psi_{1R}\\ \psi_{2R}\\ \psi_{3R}\end{pmatrix} \equiv R\Psi_R, \label{eq:neurot}
\end{equation}
where $\psi_{iR}$'s are the mass eigenstates. Without loss of generality, we take $m_{\psi_1}<m_{\psi_2}<m_{\psi_3}$. That is, $\psi_{1R}$ is the observed light neutrino. To the lowest non-trivial order in $\mu_N$, we have
\begin{eqnarray}
	M_D &=& \text{diag}\left(\frac{y^2v^2}{y^2v^2 + 2g^2v_\phi^2}\mu_N,\right.\label{eq:mass}\\
	& & \left. \frac{1}{\sqrt{2}}\sqrt{y^2v^2 + 2g^2v_\phi^2} \mp \frac12\frac{y^2v^2 + 4g^2v_\phi^2}{y^2v^2 + 2g^2v_\phi^2}\mu_N \right) +\mathcal{O}(\mu_N^2)
	\nonumber
\end{eqnarray}
Notice in the limit $\mu_N\to0$, $m_{\psi_1}\to0$ and $m_{\psi_2}=m_{\psi_3}$.
As an illustrative example, the smallest eigenvalue in the case $y=g=0.1$ is shown in Fig.~\ref{fig:neucon}.
%%%%%%%%%%%%%%%%%%%%%%%%%%%%%%%%%%%%%%%%%%%%%%%%%%%%%%%%%%%%%%%%%%%%%%%%%%%%%%%%%%%%%%%%%%%%%%%%%%%%%%%%%%%%%%%%%%%
%%%%%%%%%%%%%%%%%%%%%%%%%%%%%%%%%%%%%%%%%%%%%%%%%%%%%%%%%%%%%%%%%%%%%%%%%%%%%%%%%%%%%%%%%%%%%%%%%%%%%%%%%%%%%%%%%%%

\subsection{Neutrino couplings}
First we consider the neutrino Yukawa couplings. In the interaction basis we have
\begin{eqnarray}
	\mathcal{L}_{Y} &\supset& -\frac{y}{\sqrt{2}} h'\overline{\nu_{L}}N_{1R} - g\phi'\overline{N_{2L}}N_{1R} + \hc \nonumber\\
	&=& -\frac{y}{2\sqrt{2}}h'(\overline{\nu_L}N_{1R} + \overline{\nu_L^c}N_{1R}^c) \nonumber\\
	& &- \frac{g}{2}\phi'(\overline{N_{2L}}N_{1R}+\overline{N_{2L}^c}N_{1R}^c)+\hc\\
	&=& -\frac{y}{2\sqrt{2}}h'R_{1j}R_{2k}(\overline{\Psi^c_{R}}+\overline{\Psi_R})_j(\Psi_{R}+\Psi^c_{R})_k\nonumber \\
	& & - \frac{g}{2}\phi' R_{2k}R_{3j}(\overline{\Psi^c_{R}}+\overline{\Psi_R})_j(\Psi_{R}+\Psi^c_{R})_k + \hc.\nonumber
\end{eqnarray}
Introducing Majorana field $\Psi = \Psi_R + \Psi^c_R$, the above interaction can be written as
\begin{equation}
	\mathcal{L}_{Y} = -\frac{1}{2}\left[\frac{y}{\sqrt{2}}h'R_{1j}R_{2k} + g\phi' R_{2k}R_{3j}\right]\bar\Psi_j\Psi_k + \hc.
	%\mathcal{L}_{Y} = -\frac{y}{2\sqrt{2}}hR_{1j}R_{2k}\bar\Psi_j\Psi_k - \frac{g}{2}\phi R_{2k}R_{3j}\bar\Psi_j\Psi_k + \hc
\end{equation}
Note in the above equation, $h'$ and $\phi'$ are not in the mass basis. They can be rotated to the mass basis by an orthogonal rotation, see Eq.~\eqref{eq:scalarmixing}.
%Since the scalar also mixes, we introduce the rotation to physical bases of the scalar field as
%\begin{equation}
%	\begin{pmatrix}h\\\phi\end{pmatrix} = \begin{pmatrix}c_\theta &-s_\theta\\ s_\theta & \phantom{-}c_\theta\end{pmatrix}
%		\begin{pmatrix}h'\\\phi'\end{pmatrix}.
%\end{equation}
%Thus in terms of physical fields, we get
%\begin{equation}
%\begin{aligned}
%	\mathcal{L}_{Y} &= -\frac12\left[\frac{y c_\theta}{\sqrt{2}}R_{1j}R_{2k} + g s_\theta R_{2k}R_{3j}\right] \bar\Psi_j\Psi_k h' +\hc \\
%	&\quad +\frac{1}{2}\left[\frac{y s_\theta}{\sqrt{2}}R_{1j}R_{2k} - gc_\theta R_{2k}R_{3j}\right]\bar\Psi_j\Psi_k \phi' + \hc
%\end{aligned}
%\end{equation}

Now we consider the couplings of neutrino with gauge bosons.  They arise from the kinetic term of the lepton doublet, $\ell$,
\begin{eqnarray}
	i\bar\ell\slashed{D}\ell &\supset& i\bar\ell\left(-i\frac{e}{\sqrt{2}s_{\theta_W}}(\slashed{W}^+\sigma^+ + \slashed{W}^-\sigma^-)\right.\nonumber \\
	& & \qquad\left.- i\frac{e}{2s_{\theta_W}c_{\theta_W}}\slashed{Z}(c^2_{\theta_W} \sigma^3 + s^2_{\theta_W})\right)\ell\\
	&\supset& \frac{e}{2s_{\theta_W}}\left[\sqrt{2}(\overline{e_L}\slashed{W}^-\nu_L + \overline{\nu_L}\slashed{W}^+e_L) + \frac{1}{c_{\theta_W}}\overline{\nu_L}\slashed{Z}\nu_L\right],\nonumber
\end{eqnarray}
where $c_{\theta_W}(s_{\theta_W})$ is the cosine (sine) of the Weinberg angle, $\sigma^\pm=(\sigma^1\pm i\sigma^2)/2$ and $\sigma^i$'s are the Pauli matrices. In term of physical basis, we have
\begin{eqnarray}
	\mathcal{L}_{gauge} &\supset& \frac{e}{\sqrt{2}s_{\theta_W}}R_{1j}(\overline{e_L}\slashed{W}^-P_L\Psi_j + \overline{\Psi_j}\slashed{W}^+P_Le_L) \nonumber \\
	& & + \frac{1}{2}\frac{e}{c_{\theta_w}s_{\theta_w}}R_{1j}R_{1k}\overline{\Psi_k}\slashed{Z}P_L\Psi_j.
	%&= \frac{g}{\sqrt{2}}R_{1i}(\overline{e_L}\slashed{W}^-\psi_i^c + \overline{\psi^c_i}\slashed{W}^+e_L) - \frac{1}{2}\frac{g}{c_{\theta_w}}R_{1i}R_{1j}\overline{\psi_i}\slashed{Z}\psi_j,
	\label{eq:nugauge}
\end{eqnarray}
%where we express the coupling in term of the $e$ and Weinberg angle $\theta_w$ to avoid using too many couplings $g$'s.
%%%%%%%%%%%%%%%%%%%%%%%%%%%%%%%%%%%%%%%%%%%%%%%%%%%%%%%%%%%%%%%%%%%%%%%%%%%%%%%%%%%%%%%%%%%%%%%%%%%%%%%%%%%%%%%%%%%
%%%%%%%%%%%%%%%%%%%%%%%%%%%%%%%%%%%%%%%%%%%%%%%%%%%%%%%%%%%%%%%%%%%%%%%%%%%%%%%%%%%%%%%%%%%%%%%%%%%%%%%%%%%%%%%%%%%
\subsection{Scalar Mixing}
Due to the scalar potential in Eq.~(\ref{eq_neu2}), the field $h'$ and $\phi'$ are allowed to mix. The mass matrix, in the basis ($h'$, $\phi'$) is
\begin{eqnarray}
M^2 = \begin{pmatrix} \lambda v^2 & \lambda_{\phi H}vv_{\phi} \\ \lambda_{\phi H}vv_{\phi} & \lambda_{\phi} v_{\phi}^2
\end{pmatrix}.%\begin{pmatrix} 2\lambda v^2 & \lambda_{\phi H}vv_{\phi} \\ \lambda_{\phi H}vv_{\phi} & 2\lambda_{\phi} v^2\end{pmatrix}
\end{eqnarray}
%where the basis for scalar fields is $S = (\widetilde{h^0}, \widetilde{\phi})$. 
We can diagonalize the mass matrix by an orthogonal rotation to the physical basis
\beqa
\begin{pmatrix} h \\ \phi \end{pmatrix} = \begin{pmatrix} \cos{\theta} & -\sin{\theta} \\ \sin{\theta} & \cos{\theta} \end{pmatrix} \begin{pmatrix} h' \\ \phi' \end{pmatrix}
\label{eq:scalarmixing}
\eeqa
where the mixing angle is determined by
\beq
\tan{2\theta} = \frac{\lambda_{\phi H} v v_{\phi}}{\lambda_{\phi}v_{\phi}^2 - \lambda v^2 }.
\eeq
The masses of the two physical states are
\beq
m_{h, \phi}^2 = \lambda v^2 + \lambda_{\phi}v_{\phi}^2 \mp \sqrt{(\lambda_{\phi}v_{\phi}^2 - \lambda v^2 )^2 + \lambda_{\phi H}^2 v^2 v_{\phi}^2}.\label{eq:scalarmass}
\eeq
%The physical states are related to the original states by the rotation matrix

The mixing angle $\theta$ is constrained by the LHC 125 GeV Higgs measurements. So far, the measurements have been consistent with the SM predictions~\cite{Khachatryan:2016vau,Aaboud:2018xdt,Sirunyan:2018koj}. Thus we expect the mixing angle $\theta$ to be small. 

Finally, for later conveniences, we give expressions for the parameters in the scalar potential in terms of physical masses, vevs and mixing angle% in the parameter scan, we recast $\lambda,\lambda_{\phi},\lambda_{\phi H}$ as
\begin{align}
	\lambda &= \frac{c_\theta^2m_h^2 + s_\theta^2m_\phi^2}{v^2},\\
	\lambda_\phi &= \frac{s_\theta^2m_h^2 + c_\theta^2m_\phi^2}{v_{\phi}^2},\\
	\lambda_{\phi H} &= \frac{c_\theta s_\theta (m_\phi^2-m_h^2)}{vv_{\phi}}. \label{eq:lambdascalar}
\end{align}
The scalar potential is bounded from below provided $\lambda_{\phi H} > 0$, therefore we will strictly work in the case where $m_{\phi} > m_h$.
%%%%%%%%%%%%%%%%%%%%%%%%%%%%%%%%%%%%%%%%%%%%%%%%%%%%%%%%%%%%%%%%%%%%%%%%%%%%%%%%%%%%%%%%%%%%%%%%%%%%%%%%%%%%%%%%%%%
%%%%%%%%%%%%%%%%%%%%%%%%%%%%%%%%%%%%%%%%%%%%%%%%%%%%%%%%%%%%%%%%%%%%%%%%%%%%%%%%%%%%%%%%%%%%%%%%%%%%%%%%%%%%%%%%%%%

\subsection{Dark Matter}

The coupling of $\Phi$ with $\chi$ in Eq.~\eqref{eq:darkmatter} gives extra contributions to the DM mass, $m_\chi$. This can be seen by making a chiral rotation
\beq
\chi \to \chi' = e^{i\alpha\gamma_5}\chi,
\eeq
where $\alpha = \frac{1}{2}\tan^{-1}\left(\frac{v_{\phi} \widetilde{G}}{M + v_{\phi} G}\right)$. After chiral rotation, the mass of $\chi$ is 
\beq
m_{\chi} = \sqrt{\frac{\left( \sqrt{2}M + v_{\phi} G\right)^2 + \left( v_{\phi} \widetilde{G}\right)^2}{2}} \label{eq:dmmass}.
\eeq
The interaction Lagrangian becomes
\beq
\mathcal{L} = \frac{v_{\phi} \left( G^2 + \widetilde{G}^2\right)}{m_{
\chi}}\phi\overline{\chi'}\chi' + \frac{M}{m_{\chi}}\phi \overline{\chi'}\left( G + i\widetilde{G}\gamma_5 \right)\chi'
\eeq
Notice that if $M = 0$ the interaction Lagrangian would contain no pseudoscalar coupling after the chiral transformation. Where the limit $v_{\phi} = 0$ leads to no change in the Lagrangian.
%%%%%%%%%%%%%%%%%%%%%%%%%%%%%%%%%%%%%%%%%%%%%%%%%%%%%%%%%%%%%%%%%%%%%%%%%%%%%%%%%%%%%%%%%%%%%%%%%%%%%%%%%%%%%%%%%%%
%%%%%%%%%%%%%%%%%%%%%%%%%%%%%%%%%%%%%%%%%%%%%%%%%%%%%%%%%%%%%%%%%%%%%%%%%%%%%%%%%%%%%%%%%%%%%%%%%%%%%%%%%%%%%%%%%%%
%%%%%%%%%%%%%%%%%%%%%%%%%%%%%%%%%%%%%%%%%%%%%%%%%%%%%%%%%%%%%%%%%%%%%%%%%%%%%%%%%%%%%%%%%%%%%%%%%%%%%%%%%%%%%%%%%%%

\section{Phenomenology}\label{ch:pheno}
%%%%%%%%%%%%%%%%%%%%%%%%%%%%%%%%%%%%%%%%%%%%%%%%%%%%%%%%%%%%%%%%%%%%%%%%%%%%%%%%%%%%%%%%%%%%%%%%%%%%%%%%%%%%%%%%%%%
\subsection{Invisible decay of $Z$}
The mixing between the neutrinos impacts the coupling of light neutrinos to $Z$ boson, see Eq.~\eqref{eq:nugauge}. This results in a modification to the partial decay width of the $Z$ boson into neutrinos. %and thus it is constrained by the invisible decay width of the $Z$ boson.
The invisible decay width of the $Z$ boson has been measured very precisely, $\Gamma^{exp}_{inv} = 499.0\pm1.5$ MeV, while the SM prediction is $\Gamma_{inv}^{\text{SM}} = 501.66\pm0.05$ MeV \cite{Tanabashi:2018oca}. If we assume only $Z\to\psi_1\psi_1$ is kinematically allowed, we would get
\begin{equation}
	\Gamma_{inv} = R_{11}^4\Gamma_{inv}^{\text{SM}},
\end{equation}
where $R_{11}$ is the 1-1 component of the rotation matrix defined in Eq.~(\ref{eq:neurot}). This places a $2\sigma$ limit on $R_{11}$ as
\begin{equation}
	R_{11}^4 \gtrsim \frac{496}{501} \Longrightarrow R_{11} \gtrsim 0.997.
\end{equation} 
It translates, in terms of Lagrangian parameters, to the bound 
\begin{equation}
	\frac{v_\phi}{\sqrt{\left(\frac{y}{\sqrt{2}g}\right)^2v^2 + v^2_\phi}} \gtrsim 0.997. \label{eq:zinvis}
\end{equation}
To get a feel for this constraint, let's take $v = 246$ GeV and $v_\phi=1000$ GeV, we get $y/g \lesssim 0.43$.

%%%%%%%%%%%%%%%%%%%%%%%%%%%%%%%%%%%%%%%%%%%%%%%%%%%%%%%%%%%%%%%%%%%%%%%%%%%%%%%%%%%
\subsection{125 GeV Higgs Data}
The mixing angle $\theta$ changes the coupling of the 125 GeV Higgs boson to other SM particles. These couplings have been measured to about 10\% accuracy at the LHC. All the measurements can be parametrized in term of a coupling strength modifier
\begin{equation}
	\mu_i^f = \frac{\sigma_i}{\sigma_i^{(SM)}}\frac{Br^f}{Br^f_{(SM)}},
\end{equation} 
where $i$ indicates the production cross-section channel of the 125 Higgs boson, $f$ indicates the branching ratio channel. Using both LHC Run 1~\cite{Khachatryan:2016vau} and Run 2~\cite{Aaboud:2018xdt,Sirunyan:2018koj} data, we deduce the overall best fit value for $\mu = 1.09\pm0.07$.

In our model, all the Higgs measurements are modified by the mixing angle, $\cos\theta$. Thus the predicted value of $\mu$ is $\hat\mu=\cos^4\theta$. Therefore, consistency with the Higgs data requires $|\cos\theta|\ge0.9931$ at 95\% confidence levels.

%%%%%%%%%%%%%%%%%%%%%%%%%%%%%%%%%%%%%%%%%%%%%%%%%%%%%%%%%%%%%%%%%%%%%%%%%%%%%%%%%%%%%%%%%%%%%%%%%%%%%%%%%%%%%%%%%%%

\subsection{Indirect detection}
DM $\chi$ can self-annihilate through their interaction with $\Phi$, see Eq.~\eqref{eq:darkmatter}. We give explicit expressions for all the 2-2 annihilation channels of $\chi$ in App.~\ref{sec:annihilation}. From there, we see that the annihilations into the SM gauge bosons and fermions are suppressed by the scalar mixing angle. Thus the main annihilation channels for $\chi$ are $\chi\overline{\chi} \rightarrow \phi \phi, \phi h, h h$ and $\chi\overline{\chi} \rightarrow \psi\bar\psi$. 
The annihilation into neutrino can be looked for at neutrino telescopes such as IceCUBE~\cite{Aartsen:2017ulx}. For annihilations into $h$ and $\phi$, they can subsequently decays into photons which can be looked for with gamma ray telescopes such as Fermi-LAT~\cite{TheFermi-LAT:2017vmf}.

%Typically, the main annihilation channels are $\chi\overline{\chi} \rightarrow \phi \phi, \phi h, h h$ and $\chi\overline{\chi} \rightarrow \nu \nu$ as shown in Fig. (\ref{fig:xsec}). The cascade decay such as gamma ray and neutrinos from dark matter annihilation at the galactic centers or the center of the Sun could provide a unique signal for astrophysics experiments. For example, the dark matter annihilation into scalar particles then lead to an indirect detection through gamma ray measurement such as Fermi satellite telescope while the neutrino products from dark matter annihilation could be detected from the neutrino telescope experiment.
\subsubsection{Gamma ray}
The gamma ray flux produced from DM annihilation is given by 
\beq
\frac{d\Phi}{dE_{\gamma}} = \frac{J R_{\rm sc} \rho_{\rm sc}^2}{8\pi m_{\rm dm}^2} \sum_{i,j}  \langle \sigma v \rangle_i\text{Br}_{ij} \frac{dN_j}{dE_{\gamma}} \label{eq:flux}
\eeq
where $i$ runs over different scalar annihilation channels, $j$ runs over different SM final states, Br$_{ij}$ is the branching ratio from initial state $i$ into SM final state $j$, and $\frac{dN_j}{dE_{\gamma}}$ gives the gamma ray spectrum from the SM particle $j$. $R_{\rm sc} = 8.5$ kpc and $\rho_{\rm sc}$ are normalization constants introduced to make $J$ dimensionless.
Here, $R_{\rm sc}$ is the distance between the Sun and Milky way's center, $\rho_{\rm sc}$ is the DM density at position of the Sun. The $J$ factor is the typical average line of sight integral over the DM halo
\begin{equation}
%J = \frac{1}{2R_{\rm sc} \rho_{\rm sc}^2}\int\displaylimits_{-1}^{1}d\cos\theta\int\displaylimits_0^{l_{\rm max}}dl\, \rho^2\left(\sqrt{R_{\rm sc}^2 -2lR_{\rm sc}\cos\theta + l^2}\right),
	J = \frac{1}{2R_{\rm sc} \rho_{\rm sc}^2}\int\displaylimits_{-1}^{1}d\cos\theta\int\displaylimits_0^{l_{\rm max}}dl\, \rho^2(x),
\end{equation}
where $x=\sqrt{R_{\rm sc}^2 -2lR_{\rm sc}\cos\theta + l^2}$ is the distance from the galactic center to the position along the line of sight and $l_{\rm max} = \sqrt{R_{\rm halo}^2 - R_{\rm sc}^2 \sin^2\theta} + R_{\rm sc} \cos\theta$ is the distance along the line of sight to the edge of the galaxy.
%The dark matter profile is assumed to be the NWF profile written as
%\beq
%\rho(r) = \frac{\rho_0}{\left(\frac{r}{r_s}\right)\left( 1+ \frac{r}{r_s}\right)^2}
%\eeq
%where $r_s = 16.1, \rho_{\rm sc} = 0.471$.
The integral over the line of sight gives the value of $J=3.34$ for the NFW profile and $J=1.60$ for the Burkert profile. 
In our analysis, we use the NFW profile. One can easily translate our result to other DM profile by an appropriate rescaling of the $J$ factor.%, the reader can rescale our result twe decide to use a stringent limit so only NFW profile will be considered in this work.

The gamma ray spectrum coming from the charged-particle final states can be obtained through computer simulations. Since we only need a ballpark estimation in order to obtain the constraint, the spectrum is assumed to have a power-law relation.
However, in our case, the SM particles $j = {b, t, u, W^{\pm}/Z}$ are produced from the subsequent decay of scalar particles $\chi \overline{\chi} \rightarrow \phi \phi, h h, \phi h \rightarrow j\overline{j} j' \overline{j'}$. In the rest frame of the scalar $\phi$, (or $h$), the 4-momentum of the final state particle are isotropically distributed. Boosting back to the DM center of mass frame, the energy of the final states $E_j$ ranges between
\begin{align*}
E_{\rm min}^{\phi,h} &= \frac{m_{\chi}}{2}\left( 1- \sqrt{1-\left(\frac{m_{\phi,h}}{m_{\chi}}\right)^2}\sqrt{1-\left(\frac{2m_{j}}{m_{\phi,h}}\right)^2} \right)\\
E_{\rm max}^{\phi,h} &= \frac{m_{\chi}}{2}\left( 1 +  \sqrt{1-\left(\frac{m_{\phi,h}}{m_{\chi}}\right)^2}\sqrt{1-\left(\frac{2m_{j}}{m_{\phi,h}}\right)^2} \right).
\end{align*}
Averaging all possible direction, the differential probability of finding the SM particle with energy $E_{j}$ is
\beq
%\left(\frac{dP}{dE_{j}}\right)_{\phi,h} = \frac{4}{\pi m_{\chi}}\frac{\sqrt{\left(1-\frac{m_{\phi,h}^2}{m_{\chi}^2}\right)\left(1-\frac{4 m_{j}^2}{m_{\phi,h}^2}\right) - \left( 1-\frac{2E_{j}}{m_{\chi}}\right)^2}}{\left(1-\frac{m_{\phi,h}^2}{m_{\chi}^2}\right)\left(1-\frac{4 m_{i}^2}{m_{\phi,h}^2}\right)}
\begin{aligned}
\left(\frac{dP}{dE_{j}}\right)_{\phi,h} &= \frac{4}{\pi m_{\chi}}\frac{1}{\left(1-\frac{m_{\phi,h}^2}{m_{\chi}^2}\right)\left(1-\frac{4 m_{j}^2}{m_{\phi,h}^2}\right)}\\
&\quad\times\scriptstyle\sqrt{\left(1-\frac{m_{\phi,h}^2}{m_{\chi}^2}\right)\left(1-\frac{4 m_{j}^2}{m_{\phi,h}^2}\right) - \left( 1-\frac{2E_{j}}{m_{\chi}}\right)^2}.
\end{aligned}
\eeq
%This means that the gamma ray spectrum must be convoluted with the above probability distribution which is written as
Finally, the gamma ray spectrum in the center of mass frame of the annihilating DM can be written as
\begin{eqnarray}
\label{eq:gammaflux}
\left(\frac{dN_{j}}{dE_{\gamma}}\right)_{\phi\phi} &=& \int_{E_{\rm min}^{\phi}}^{E_{\rm max}^{\phi}} dE_{j} \frac{a_j E_{j}^{1/2} }{E_{\gamma}^{3/2}} e^{-b_j E_{\gamma}/E_{j}}2\left(\frac{dP}{dE_{j}}\right)_{\phi},\nonumber \\
\left(\frac{dN_{j}}{dE_{\gamma}}\right)_{h h} &=& \int_{E_{\rm min}^{h}}^{E_{\rm max}^{h}} dE_{j} \frac{a_j E_{j}^{1/2} }{E_{\gamma}^{3/2}} e^{-b_j E_{\gamma}/E_{j}}2\left(\frac{dP}{dE_{j}}\right)_{h},\nonumber \\
\left(\frac{dN_{j}}{dE_{\gamma}}\right)_{\phi h} &=& \int_{E_{\rm min}^{\phi}}^{E_{\rm max}^{\phi}} dE_{j} \frac{a_j E_{j}^{1/2} }{E_{\gamma}^{3/2}} e^{-b_j E_{\gamma}/E_{j}} \\
& & \times \left( \left(\frac{dP}{dE_{j}}\right)_{\phi} + \left(\frac{dP}{dE_{j}}\right)_{h} \right).\nonumber 
\end{eqnarray}
where $(a_j,b_j)$ are power law parameters which their values can be obtained from the simulation \cite{Bergstrom:1997fj,Feng:2000zu}. The values of the parameters are shown to be 
\beq
(a_j,b_j) = (1.0,10.7), (1.1,15.1),(0.95,6.5),(0.73,7.76).\nonumber
\eeq
for bottom quarks, top quarks, up quarks and gauge bosons final states.

In the case of light mediator $m_{\phi} < m_{j}$ and $m_{\phi} < 2m_h$ where it cannot decay into particles outside the SM, the branching ratios of $\phi$ is similar to those of Higgs since it decays through mixing with the Higgs. If the channel of sterile neutrino final states are open, $m_{\phi} > m_{j}$, or the 2 Higgs final states are open, $m_{\phi} > 2m_h$ the branching ratio into the 2 SM particles is shown in App.~\ref{ch:decaywidth}. Although sterile neutrinos are unstable and subsequently decay into SM particles via off-shell $\phi$, such decay involves 3-particle final states where at least one of them is neutrino, $\phi \rightarrow \nu + {\rm SM + SM}$. Similarly, the 2 Higgs final states subsequently decay into multiple SM particles. The electromagnetic showering energy from these channels are therefore assumed to be subdominant and irrelevant to our study.
We will further approximate that for each mass range of the scalar particle, the decay is 100\% into the largest contribution to the decay width.
Therefore, the power spectrum is chosen  according to the final states as shown below:
\begin{eqnarray}
\label{eq:powerlawindex}
(a,b) = \begin{cases} (0.73,7.76),\;\; 160 {\rm\;\;GeV} \leq m_{\phi},\\ (1.0,10.7),\;\; 125 {\rm \;\;GeV} < m_{\phi} <160 {\rm \;\;GeV}.
%9 {\rm \;\;GeV} < m_{\phi} <160 {\rm \;\;GeV}\\ (0.95,6.5),\;\; m_{\phi} \leq 9 {\rm \;\;GeV} 
\end{cases}
\end{eqnarray}

\begin{figure}
\centering
\includegraphics[width=0.95\linewidth]{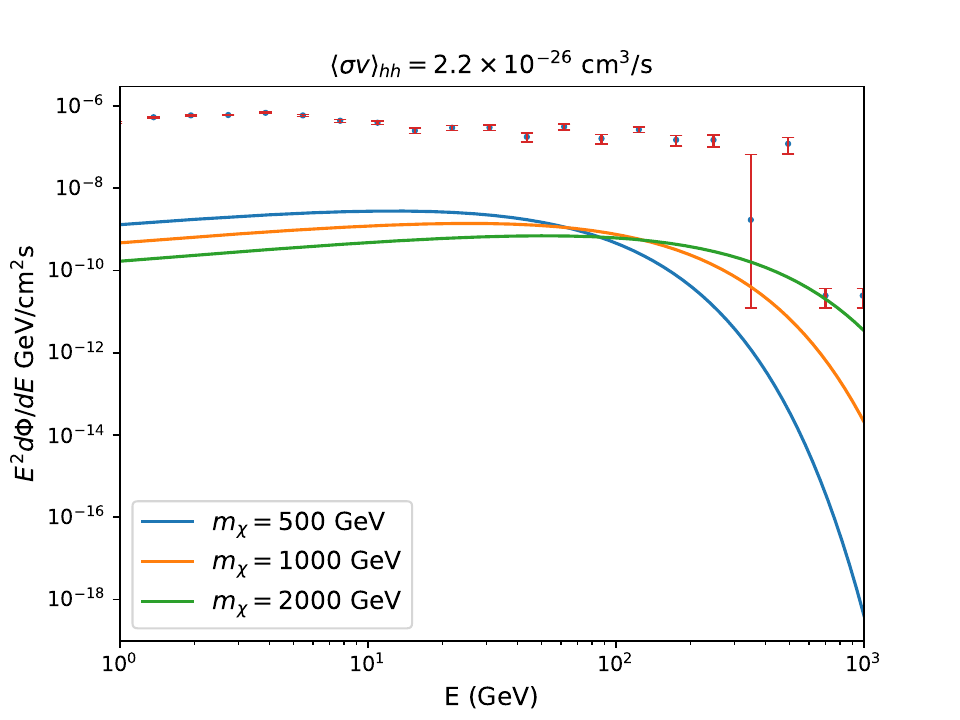}
\includegraphics[width=0.95\linewidth]{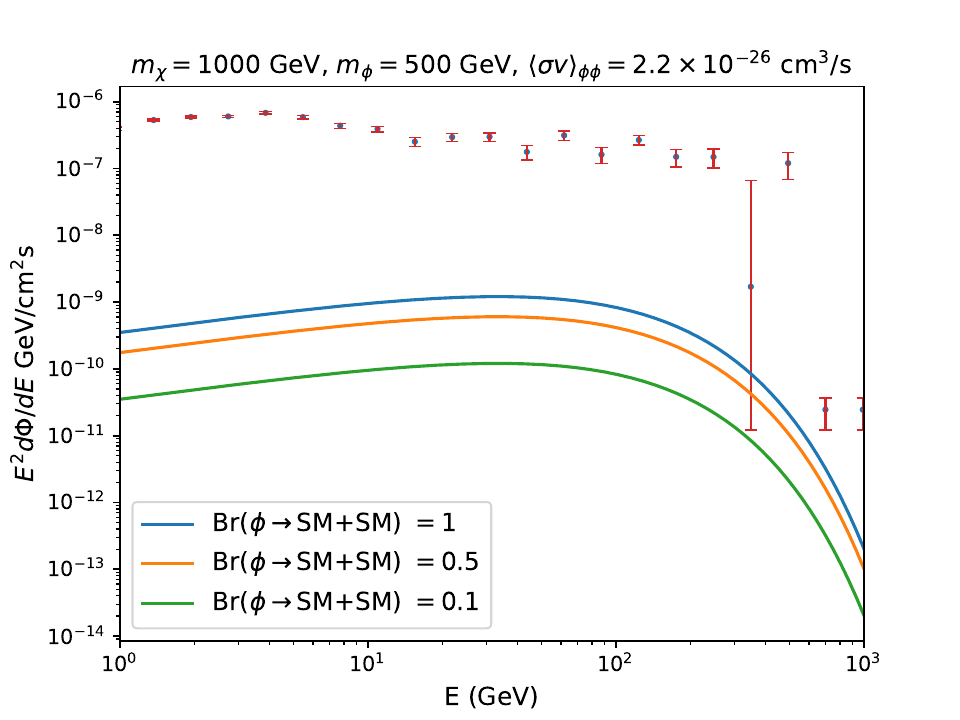}
\caption{Gamma ray flux from DM annihilation into $h h$ (above) and $\phi\phi$ (below).}
\label{fig:fluxphi}
\end{figure}

As an illustrative example, let us investigate the gamma ray flux from DM annihilation in Eq.~\eqref{eq:gammaflux}. For simplicity, we consider then extreme case where DM annihilate through a single channel and its annihilation cross-section is equal to the thermal relic value. In the case where DM annihilate into a pair of $h$, the gamma ray flux is shown in the top pane of Fig.~\ref{fig:fluxphi}. The bottom panel shows the gamma ray flux in the case where DM annihilate into $\phi\phi$ with $m_\phi=500$ GeV. We see that for these two extreme cases, the fluxes differ by less than an order of magnitude. However, in computing the gamma ray flux for a particular model parameter point, collectively represented by ($\langle \sigma v \rangle_{\rm scalar}, m_{\chi}, m_{\phi}$), one needs to take into account annihilation through all three channels in Eq.~\eqref{eq:gammaflux}. We perform such a computation in Sec.~\ref{ch:results}.

%Fig.~\ref{fig:fluxphi} shows gamma ray flux obtained from Eq.~\eqref{eq:gammaflux} for two different DM annihilation channels. In the top figure, the flux is obtained assuming DM annihilate only through $\chi\chi\to hh$ channel with the annihilation cross-section equals to the thermal relic cross-section. The bottom figure assumes that DM annihilate only through $\chi\chi\to \phi\phi$ with the annihilation cross-section again equals to the thermal relic value. We have also assumed $m_\phi=500 GeV$ for the bottom figure.

%The above equation and the flux equation (\ref{eq:flux}) are used to calculate the flux from each model point ($\langle \sigma v \rangle_{\rm scalar}, m_{\chi}, m_{\phi}$). The spectrum is shown in Fig.~\ref{fig:fluxphi} where the value of the flux can be seen to only vary less than an order of magnitude.

Let us end this subsection with a remark regarding the Fermi excess. Due to the high tail behaviour of the excess from Fermi telescope, the flux from DM annihilation cannot possibly explain all bins of the excess. It is possible some part of the energy range might originate from other astrophysical sources. In this project we will use the excess as the upper limit on the model-generating flux.

%%%%%%%%%%%%%%%%%%%%%%%%%%%%%%%%%%%%%%%%%%%%%%%%%%%%%%%%%%%%%%%%%%%%%%%%%%%%%%%%%%%%%%%%%%%%%%%%%%%%%%%%%%%%%%%%%%%

\subsubsection{Neutrino telescope}
The neutrino produced from DM annihilation carries a definite energy depending on the decay channel
\beq
E_{\nu} = \begin{cases}
m_\chi &\chi\overline{\chi} \rightarrow \nu \nu\\
\frac{4m_\chi^2 - m_{\psi}^2}{4m_\chi} &\chi\overline{\chi} \rightarrow \nu \psi_2,\nu \psi_3
\end{cases},
\eeq
where we have used $\nu\equiv\psi_1$ for the lightest neutrino.  
The flux is therefore written as
\beqa
\frac{d\Phi}{dE_{\nu}} &=& \frac{1}{3}\frac{J R_{\rm sc} \rho_{\rm sc}^2}{8\pi m_{\chi}^2}  \left[ \langle \sigma v \rangle_{\nu \nu} \delta(E_{\nu} - m_\chi)\phantom{\frac{m_{\psi_i}^2}{4m_f}}\right. \\
& & + \left. \sum_{i = 2}^3 \langle \sigma v \rangle_{\nu \psi_i} \delta\left(E_{\nu} - \frac{4m_\chi^2 - m_{\psi_i}^2}{4m_\chi}\right) \right]\nonumber,
\eeqa
where we have focussed on the electron-neutrino flux only. The 1/3 factor in the above equation arises from our assumption that $\nu$ at production and neutrino oscillation along the way from the production point to Earth result in equal ratio for each neutrino flavor.  

In our analysis, we have reinterpreted the IceCUBE data for DM search in our scenario. It turns out the annihilation cross-section into neutrinos from our model are of order $\lesssim 10^{-31}$ cm$^3$/s which is well below the recast IceCUBE limit ($\sim 10^{-24}$ cm$^3$/s). %\cite{Aartsen:2017ulx}. %This is due to the fact that  primililary 

%{\color{red} Say something about the constraints being too weak for us to consider}

%%%%%%%%%%%%%%%%%%%%%%%%%%%%%%%%%%%%%%%%%%%%%%%%%%%%%%%%%%%%%%%%%%%%%%%%%%%%%%%%%%%%%%%%%%%%%%%%%%%%%%%%%%%%%%%%%%%

\subsection{Direct detection}
In the direct detection experiment, the momentum exchanged between DM and nucleon is typically much smaller than the mass of the scalar mediator. Thus it is convenient to describe DM-nucleon interaction with an the effective operator. In our model, the effective operator for DM-nucleon interaction reads
\begin{equation}
	\mathcal{L}_{\chi N} = s_{2\theta}\left(\frac{1}{m_\phi^2}-\frac{1}{m_h^2}\right)\bar \chi(G+i\tilde{G}\gamma_5) \chi\, \mathcal{S}_q,
%	\left[ \sum_q \frac{m_q}{v}\bar q q - \frac{\alpha_s}{4\pi v}G^a_{\mu\nu}G^{a\mu\nu}\right].
\end{equation}
where $\mathcal{S}_q$ is the scalar current representing the interaction between the mediator and the quarks inside the nucleon. In the case that the momentum exchanged are smaller than the heavy quarks, the scalar current is given by 
\begin{equation}
	\mathcal{S}_q = \sum_{q=u,d,s} \frac{m_q}{v}\bar q q - \frac{\alpha_s}{4\pi v}G^a_{\mu\nu}G^{a\mu\nu},
\end{equation}   
%where the sum runs over light quark $q=u,d,s$ and 
where the gluonic term arises from integrating out the heavy quarks. 
In the case of heavy DM ($m_\chi\ge1$ TeV), the momentum exchanged can be comparable to the charm mass. For such a case, one need to take into account the charm mass threshold effect. However, we will ignore the charm mass effect in the rest of this work.

The amplitude for DM-nucleon scattering depends on the $\mathcal{S}_q$ nucleon matrix elements. They are conventionally parametized in terms of the quark and the gluonic form factors~\cite{Jungman:1995df}
\begin{equation}
\begin{aligned}
	\langle N|m_q\bar q q|N\rangle &= m_Nf_{Tq}^{(N)},\\
	\langle N|\frac{\alpha_s}{4\pi}G^a_{\mu\nu}G^{a\mu\nu}|N\rangle &= -\frac29m_Nf_{TG}^{(N)}.
\end{aligned}
\end{equation}
The gluonic form factor is related to the quark form factors by the QCD trace anomaly in the heavy quark limit~\cite{Shifman:1978zn}
\begin{equation}
	f_{TG}^{(N)} = 1 - \sum_{q=u,d,s}f_{Tq}^{(N)}.
\end{equation}
Thus the nucleon matrix element of the scalar current is
\begin{equation}
	f_N \equiv \langle N|S_q|N\rangle = \frac29\frac{m_N}{v}\left(1+\frac72\sum_{q=u,d,s}f_{Tq}^{(N)}\right).
	\label{eq:matrixelement}
\end{equation}

For numerical analysis, we take the strange quark form factor to be $f_{Ts}^{(p)} = f_{Ts}^{(n)} = 0.043\pm0.011$~\cite{Junnarkar:2013ac}. We extract the up and the down quark form factors from the pion-nucleon sigma term, $\sigma_{\pi N}$, using the relations provided by Ref.~\cite{Crivellin:2013ipa}. However, there is a discrepancy between the values of $\sigma_{\pi N}$ extracted from the scattering data using baryon chiral effective theory and the lattice computation. We follow Ref.~\cite{Bishara:2015cha} and conservatively taking $\sigma_{\pi N}=50\pm15$ MeV.  Thus we determine the $u$ and $d$ quark form factors to be
\begin{equation}
\begin{aligned}
	f^{(p)}_{Tu} &= (1.8\pm0.5)\times10^{-2},\\
	f^{(p)}_{Td} &= (3.4\pm1.1)\times10^{-2},
\end{aligned}
\qquad
\begin{aligned}
	f^{(n)}_{Tu} &= (1.6\pm0.5)\times10^{-2},\\
	f^{(n)}_{Td} &= (3.8\pm1.1)\times10^{-2}.
\end{aligned}
\end{equation}
Therefore, to a good approximation, the nucleon matrix elements, Eq.~\eqref{eq:matrixelement}, for the proton and the neutron  are the same $\langle p|S|p\rangle \simeq \langle n|S|n\rangle$. 

Armed with the DM-nucleon matrix element, we determine the spin-average DM-nucleon scattering amplitude squared, in the zero momentum transferred limit, to be
\begin{equation}
	|\mathcal{M}|^2 = 4f_N^2(G^2+\tilde G^2)s_{2\theta}^2m_N^2m_\chi^2\left(\frac{1}{m_\phi^2}-\frac{1}{m_h^2}\right)^2.
\end{equation}
Finally, we determine the DM-nucleon scattering cross-section to be 
\begin{equation}
	\sigma_{\chi N} = \frac{f_N^2(G^2+\tilde G^2)s_{2\theta}^2}{4\pi}\frac{m_\chi^2m_N^2}{(m_\chi+m_N)^2}\left(\frac{1}{m_\phi^2}-\frac{1}{m_h^2}\right)^2. \label{eq:ddxsec}
\end{equation}
The current upper limit is reported by XENON1T collaboration \cite{Aprile:2017iyp}.
%%%%%%%%%%%%%%%%%%%%%%%%%%%%%%%%%%%%%%%%%%%%%%%%%%%%%%%%%%%%%%%%%%%%%%%%%%%%%%%%%%%%%%%%%%%%%%%%%%%%%%%%%%%%%%%%%%%
%%%%%%%%%%%%%%%%%%%%%%%%%%%%%%%%%%%%%%%%%%%%%%%%%%%%%%%%%%%%%%%%%%%%%%%%%%%%%%%%%%%%%%%%%%%%%%%%%%%%%%%%%%%%%%%%%%%
%%%%%%%%%%%%%%%%%%%%%%%%%%%%%%%%%%%%%%%%%%%%%%%%%%%%%%%%%%%%%%%%%%%%%%%%%%%%%%%%%%%%%%%%%%%%%%%%%%%%%%%%%%%%%%%%%%%

\section{Results}\label{ch:results}

We approach the phenomenology of the model by scanning parameters space subjected to all constraints. We first consider the mass scale from the following random sets
\begin{align}
M, m_{\phi}, v_{\phi}  &\in [10^2,10^6]\;\; {\rm GeV}, \, |\cos\theta| \in [0.9931,1]. %\nonumber \\
%m_{\phi} &\in [10^2,10^6]\;\; {\rm GeV} \nonumber \\
%v_{\phi} &\in [10^2,10^6]\;\; {\rm GeV}\\
%\theta &\in [0,2\pi] \nonumber
\end{align}
Then we calculate the couplings in scalar sector using Eq.~(\ref{eq:lambdascalar}).
Demanding that all couplings are perturbative, we apply the constraints on $\lambda_{\phi},\lambda_{h},\lambda_{\phi h} < 4\pi$.
%\subsection{Neutrino mixing Constraints}
The next step in generating a set of parameters is to consider the neutrino sector. The set of parameters $(y,g)$ is generated from the following range:
\begin{align}
y,g &\in [10^{-3},\sqrt{4\pi}].%\\
%g &\in [10^{-3},\sqrt{4\pi}]\nonumber
\end{align}
The constraint on Z invisible decay width from Eq.~(\ref{eq:zinvis}) is then applied to the parameter set.
Finally, the rest of parameters are chosen as follows
\begin{align}
G, \widetilde{G} &\in [10^{-3},\sqrt{4\pi}], \quad \mu_N \in [10^{-10},10^{-7}]\;\;{\rm GeV}. %\nonumber \\
%\widetilde{G} &\in [10^{-3},\sqrt{4\pi}]\\
%\mu_N &\in [10^{-10},10^{-7}]\;\;{\rm GeV} %\nonumber
\end{align}

\begin{figure}
\centering
\includegraphics[width=0.95\linewidth]{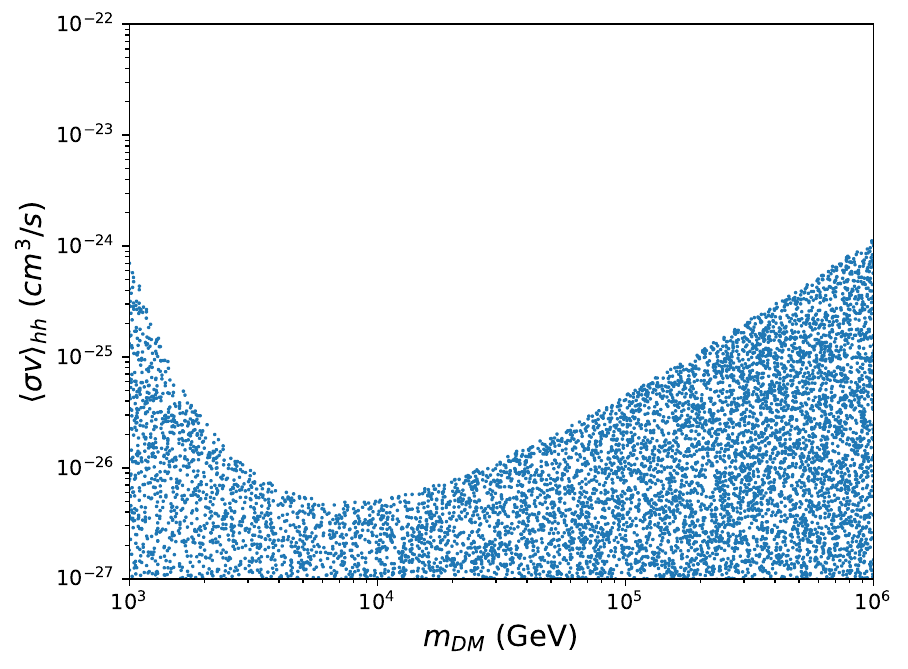}
\caption{Cross-section for DM annihilation into a pair of Higgs bosons consistent with constraints from the light neutrino mass limit, the invisible $Z$ decay width, the 125 GeV Higgs data and the Fermi gamma ray excess.}
\label{fig:gammaonhiggs}
\end{figure}

After we obtain the complete set of model parameters space, the mass spectrum of the theory is then calculated from Eq.~(\ref{eq:neumass}), (\ref{eq:scalarmass}), (\ref{eq:dmmass}). The DM annihilation cross sections are computed using the expression given in App.~\ref{sec:annihilation}. Then the neutrino mass limit is applied ($m_{\nu} < 0.2$ eV). Next we use Eq.~(\ref{eq:flux}) to produce gamma ray flux for each point of the set. 
Then we impose the excess reported by Fermi-LAT collaboration as the upper limit for the flux from DM annihilation~\cite{TheFermi-LAT:2017vmf}. Note that from Fig.~\ref{fig:gammaonhiggs}, the gamma ray constraint have a clear impact on annihilation cross section into two Higgs final states as expected since the branching ratio of the scalar mediator to the SM particles is often found too small. 

\begin{figure}
\centering
\includegraphics[width=0.95\linewidth]{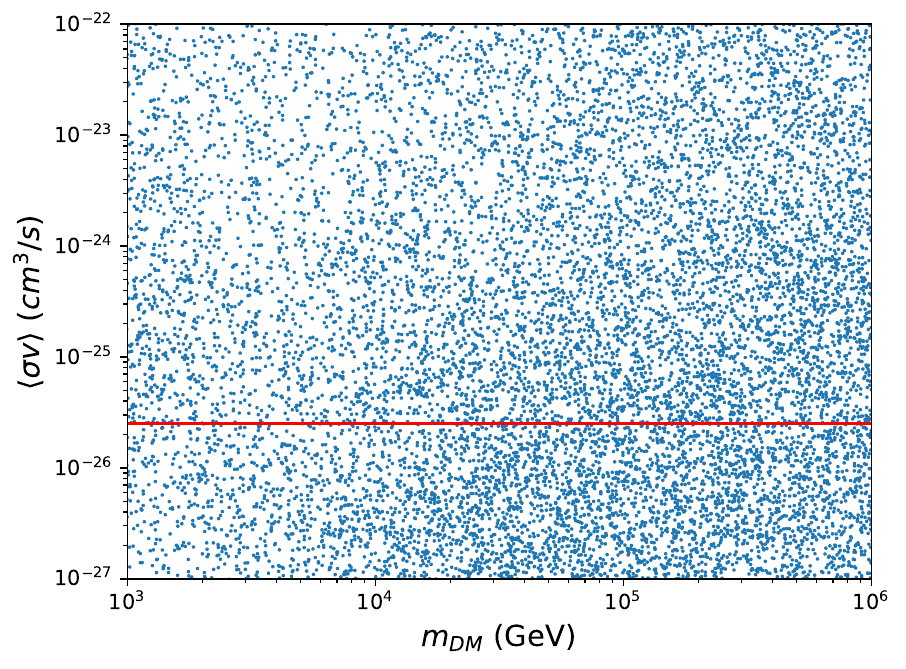}
\caption{Total DM annihilation cross-section consistent with constraints from the light neutrino mass limit, the invisible $Z$ decay width, the 125 GeV Higgs data and the Fermi gamma ray excess. The red line is the thermal relic annihilation cross-section.}
\label{fig:gammaontotal}
\end{figure}

\begin{figure}
\centering
\includegraphics[width=0.95\linewidth]{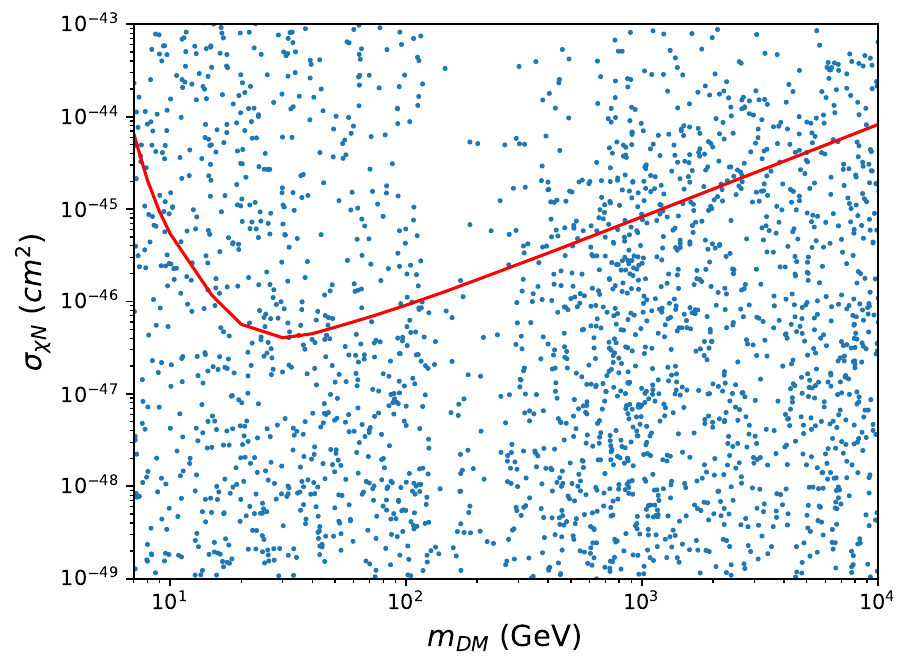}
\caption{DM-nucleon scattering cross-section consistent with constraints from the light neutrino mass limit, the invisible $Z$ decay width, the 125 GeV Higgs data, the Fermi gamma ray excess and freeze out production. The red line is the upper bound from XENON1T.} 
\label{fig:ddplot}
\end{figure}

The total DM annihilation cross-section of the remaining parameter points are shown in Fig.~\ref{fig:gammaontotal} where the thermal relic density $\langle \sigma v \rangle_{\rm total} = 2.5\times10^{-26}$ cm$^3$/s is shown as the red line. 
To prevent the universe from being over closed,
 %of the DM energy density from thermal freeze out mechanism, 
we impose the constraint $\langle \sigma v \rangle_{\rm total} > 2.5\times10^{-26}$ cm$^3$/s. Finally, we calculate the DM-nucleon scattering cross section on the remaining parameter points using Eq.~\ref{eq:ddxsec}. In Fig.~\ref{fig:ddplot}, the result is shown together with the upper limit reported by XENON1T collaboration.


\section{Summary and Outlook}\label{ch:con}

In this work, we consider a class of models in which the neutrino mass can be explained by the inverse seesaw mechanism. The smallness of the active neutrino mass is achieved through the dynamical Majorana mass of 2 additional sterile neutrinos. The extra scalar field is connected to the DM sector giving a strong connection between neutrino and DM. 

We have identified the viable parameter space of the model consistent with constraints from light neutrino mass limit, the invisible $Z$ decay width, the 125 GeV Higgs measurements and the Fermi gamma ray excess. We find that the Fermi gamma ray excess places a strong constraint on the DM annihilation to a pair of Higgs bosons. However, a large chunk of parameter space  still remains open as can be seen in Fig.~\ref{fig:ddplot}. These parts of parameter space could be probed by the next generation of direct and indirect detection experiments. It is interesting to study the sensitivities of the upcoming direct detection experiments such as LZ and XENONnT, as well as indirect detection experiment such at CTA on this model. We leave such study for possible future work.

Finally, we want to remark that our model is not yet fully realistic in the sense that it only contains 1 massive light neutrino. However, one can easily extend the model by introducing additional pair of sterile neutrinos and a flavour symmetry. We do not expect it to have a significant impact on DM phenomenology which is the main goal of the paper. 

%The generation of neutrino masses and particle spectrum of the model have been discussed. The indirect detection signal from DM annihilation have been analysed. The direct detection cross section has been calculated. We perform a numerical scan over parameters space.

\section{Acknowledgement}
The work of CP and PU has been supported in part by the Thailand Research Fund under contract no. MRG6280131 and MRG6280186, and by the National Astronomical Research Institute of Thailand. PU also acknowledges the support of the Faculty of Science, Srinakharinwirot University. NT is supported by the scholarship from the Development and Promotion of Science and Technology Talents Project (DPST). We thank the referee for his/her careful reading of our paper and for
the valuable comments and suggestions which help
improve our paper.%CP and PU have been supported by National Astronomical Research Institute of Thailand.% and the Faculty of Science, Khon Kaen University under grant no.xxx/xxxx
%%%%%%%%%%%%%%%%%%%%%%%%%%%%%%%%%%%%%%%%%%%%%%%%%%%%%%%%%%%%%%%%%%%%%%%%%%%%%%%%%%%%%%%%%%%%%%%%%%%%%%%%%%%%%%%%%%%
%%%%%%%%%%%%%%%%%%%%%%%%%%%%%%%%%%%%%%%%%%%%%%%%%%%%%%%%%%%%%%%%%%%%%%%%%%%%%%%%%%%%%%%%%%%%%%%%%%%%%%%%%%%%%%%%%%%
%%%%%%%%%%%%%%%%%%%%%%%%%%%%%%%%%%%%%%%%%%%%%%%%%%%%%%%%%%%%%%%%%%%%%%%%%%%%%%%%%%%%%%%%%%%%%%%%%%%%%%%%%%%%%%%%%%%

\appendix

\section{Annihilation cross section}
\label{sec:annihilation}
The DM annihilation cross-section into a pair of SM fermions and gauge bosons are
%The formulae for DM annihilation cross sections are:
\begin{align}
	\langle\sigma v\rangle_{W W}
	&= \frac{(G^2\beta_\chi^2+\tilde{G}^2)s^2_{2\theta}}{64\pi v^2}\left(1-r_W+\frac{3}{4}r_W^2\right)\nonumber\\
	&\quad  \times \beta_W\left(\frac{s}{s-m_h^2} - \frac{s}{s-m_\phi^2}\right)^2, \displaybreak[0]\\ 
	\langle\sigma v\rangle_{Z Z}
	&= \frac{(G^2\beta_\chi^2+\tilde{G}^2)s^2_{2\theta}}{128\pi v^2}\left(1-r_Z+\frac{3}{4}r_Z^2\right)\nonumber\\
	&\quad \times\beta_Z\left(\frac{s}{s-m_h^2} - \frac{s}{s-m_\phi^2}\right)^2 \label{eq:xsec1},\displaybreak[0]\\ 
	\langle\sigma v\rangle_{\overline{\psi}\psi}
	&= \frac{(G^2\beta_\chi^2+\tilde{G}^2)s^2_{2\theta}}{32\pi}\frac{m^2_\psi s\beta_\psi^3}{v^2}\nonumber\\
	&\quad  \times\left(\frac{1}{s-m_h^2} - \frac{1}{s-m_\phi^2}\right)^2,% \\ \displaybreak[0]
\end{align}
where $r_x=m_x^2/s$ and $\beta_\chi = \sqrt{1-\frac{4m_\chi^2}{s}}$.
The cross-section for DM annihilating into a pair of scalar bosons are given by
\begin{align}
	\langle\sigma v\rangle_{hh} 
	&= \frac{\beta_\chi^2G^2+\tilde{G}^2}{32\pi}\sqrt{1-\frac{m_h^2}{m_\chi^2}} \left(\frac{4m_\chi G s_\theta^2}{2m_\chi^2-m_h^2}\right.\nonumber\\
	& \qquad\qquad \left.-\frac{s_\theta\lambda_{hhh}}{4m_\chi^2-m_h^2} - \frac{c_\theta\lambda_{\phi hh}}{4m_\chi^2-m_\phi^2}\right)^2, \\ \displaybreak[0]
	\langle\sigma v\rangle_{\phi\phi} 
	&= \frac{\beta_\chi^2G^2+\tilde{G}^2}{32\pi}\sqrt{1-\frac{m_\phi^2}{m_\chi^2}} \left(\frac{4m_\chi G s_\theta^2}{2m_\chi^2-m_\phi^2}\right.\nonumber\\
	& \qquad\qquad \left.-\frac{s_\theta\lambda_{\phi hh}}{4m_\chi^2-m_h^2} - \frac{c_\theta\lambda_{hhh}}{4m_\chi^2-m_\phi^2}\right)^2, \\ \displaybreak[0]
%	& \quad\left.-\left(\frac{s_\theta\lambda_{h^3}}{4m_\chi^2-m_h^2} + \frac{c_\theta\lambda_{\phi h^2}}{4m_\chi^2-m_\phi^2}\right)\right]\nonumber
%	\langle\sigma v_{rel}\rangle_{hh} &= \frac{\beta_\chi^2G^2+\tilde{G}^2}{32\pi}\sqrt{1-\frac{m_h^2}{m_\chi^2}} \left[\frac{16m_\chi^2 G^2 s_\theta^4}{(2m_\chi^2-m_h^2)^2}\right.\nonumber\\
%	& \quad + \left(\frac{s_\theta\lambda_{h^3}}{4m_\chi^2-m_h^2} + \frac{c_\theta\lambda_{\phi h^2}}{4m_\chi^2-m_\phi^2}\right)^2 \\
%	& \quad\left.-\frac{8m_\chi G s_\theta^2}{2m_\chi^2-m_h^2}\left(\frac{s_\theta\lambda_{h^3}}{4m_\chi^2-m_h^2} + \frac{c_\theta\lambda_{\phi h^2}}{4m_\chi^2-m_\phi^2}\right)\right]\nonumber
	\langle\sigma v\rangle_{h\phi} 
	%&= \frac{\beta_\chi^2G^2+\tilde{G}^2}{16\pi m_\chi^2}\sqrt{\left(1+\frac{m_h^2-m_\phi^2}{4m_\chi^2}\right)^2- \frac{m_h^2}{m_\chi^2}}\nonumber\\
	&= \frac{\beta_\chi^2G^2+\tilde{G}^2}{16\pi m_\chi^2}\sqrt{\lambda\left[\frac{m_h^2}{4m_\chi^2},\frac{m_\phi^2}{4m_\chi^2}\right]}\nonumber\\
	&\quad\times\left(\frac{8m_\chi G s_\theta c_\theta}{4m_\chi^2-m_h^2-m_\phi^2}\right. \nonumber\\
	&\qquad\qquad \left.- \frac{s_\theta\lambda_{\phi hh}}{4m_\chi^2-m_h^2} - \frac{c_\theta\lambda_{h\phi\phi}}{4m_\chi^2-m_\phi^2}\right)^2,
\end{align}
where $\lambda(x,y)=1+x^2+y^2-2x-2y-2xy$ is the phase space factor and
\begin{align}
	\lambda_{h h h} &= -\frac{6 s^3_{\theta} \left(m_{\phi}^2c^2_{\theta} + m_h^2s^2_{\theta} \right)}{v_{\phi}}-\frac{6 c^3_{\theta} \left(m_{h}^2c^2_{\theta} + m_{\phi}^2s^2_{\theta} \right)}{v}\nonumber\\
	& \quad -\frac{\left(m_{\phi}^2 - m_h^2\right)c_{\theta}s_{\theta}}{vv_{\phi}}\left(3v_{\phi}c^2_{\theta}s_{\theta} + 3vc_{\theta}s^2_{\theta}\right),\\\displaybreak[0]
%	\lambda_{\phi h h} &= -\frac{6 c_{\theta}s^2_{\theta} \left(m_{\phi}^2c^2_{\theta} + m_h^2s^2_{\theta} \right)}{v_{\phi}}\nonumber \\
%	& \quad -\frac{6 c^2_{\theta}s_{\theta} \left(m_{h}^2c^2_{\theta} + m_{\phi}^2s^2_{\theta} \right)}{v}\nonumber\\
%	\lambda_{\phi h h} &= -\frac{6 c_{\theta}s^2_{\theta} \left(m_{\phi}^2c^2_{\theta} + m_h^2s^2_{\theta} \right)}{v_{\phi}}-\frac{6 c^2_{\theta}s_{\theta} \left(m_{h}^2c^2_{\theta} + m_{\phi}^2s^2_{\theta} \right)}{v}\nonumber\\
	\lambda_{\phi h h} &=-3s_{2\theta}\left[m_{\phi}^2\left(\frac{c_\theta s^2_{\theta}}{v_\phi} + \frac{c^2_{\theta}s_\theta}{v} \right) + m_{h}^2\left(\frac{c^3_{\theta}}{v_\phi} + \frac{s^3_\theta}{v} \right)\right]\nonumber\\
	& \quad -\frac{\left(m_{\phi}^2 - m_h^2\right)c_{\theta}s_{\theta}}{vv_{\phi}}\nonumber\\
	& \qquad \times \left(v_{\phi}c^3_{\theta} +2vc^2_{\theta}s_{\theta} - 2v_{\phi}c_{\theta}s^2_{\theta} - vs^3_{\theta}\right), \displaybreak[0]\\
	%\lambda_{h \phi \phi} &= -\frac{6 c^2_{\theta}s_{\theta} \left(m_{\phi}^2c^2_{\theta} + m_h^2s^2_{\theta} \right)}{v_{\phi}} -\frac{6 c_{\theta}s^2_{\theta} \left(m_{h}^2c^2_{\theta} + m_{\phi}^2s^2_{\theta} \right)}{v}\nonumber\\
	%\lambda_{h \phi \phi} &= -\frac{3s_{2\theta}}{vv_\phi}\left[m_{\phi}^2\left(vc^3_{\theta} + v_\phi s^3_\theta \right) + \frac{m_{h}^2s_{2\theta}}{2}\left(v s_{\theta} + v_{\phi}c_{\theta} \right)\right]\nonumber\\
	\lambda_{h \phi \phi} &= -3s_{2\theta}\left[m_{\phi}^2\left(\frac{c^3_{\theta}}{v_\phi} + \frac{s^3_\theta}{v} \right) + m_{h}^2\left(\frac{c_\theta s^2_{\theta}}{v_\phi} + \frac{c^2_{\theta}s_\theta}{v} \right)\right]\nonumber\\
	& \quad -\frac{\left(m_{\phi}^2 - m_h^2\right)c_{\theta}s_{\theta}}{vv_{\phi}}\nonumber\\
	& \qquad \times \left(v_{\phi}s^3_{\theta} -2vc_{\theta}s^2_{\theta} - 2v_{\phi}c^2_{\theta}s_{\theta} + vc^3_{\theta}\right), \\\displaybreak[0]
	\lambda_{\phi \phi \phi} &= -\frac{6 c^3_{\theta} \left(m_{\phi}^2c^2_{\theta} + m_h^2s^2_{\theta} \right)}{v_{\phi}}+\frac{6 s^3_{\theta} \left(m_{h}^2c^2_{\theta} + m_{\phi}^2s^2_{\theta} \right)}{v}\nonumber\\
	& \quad -\frac{\left(m_{\phi}^2 - m_h^2\right)c_{\theta}s_{\theta}}{vv_{\phi}}\left(3v_{\phi}c_{\theta}s^2_{\theta} - 3vc^2_{\theta}s_{\theta}\right).
\end{align}

Finally, the annihilation cross-sections into neutrinos are given by 
%\beqa
%%	\langle\sigma v_{rel}\rangle_{\phi\phi} &=& \frac{\beta_f^2G^2+\tilde{G}^2}{32\pi}\sqrt{1-\frac{m_\phi^2}{m_f^2}}\left[\frac{16m_f^2 G^2 c_\theta^4}{(2m_f^2-m_\phi^2)^2} \right.\nonumber\\
%%	& &+ \left(\frac{s_\theta\lambda_{h\phi^2}}{4m_f^2-m_h^2} + \frac{c_\theta\lambda_{\phi^3}}{4m_f^2-m_\phi^2}\right)^2\label{eq:xsec2}\\
%%	& &\left.-\frac{8m_f G c_\theta^2}{2m_f^2-m_\phi^2}\left(\frac{s_\theta\lambda_{h\phi^2}}{4m_f^2-m_h^2} + \frac{c_\theta\lambda_{\phi^3}}{4m_f^2-m_\phi^2}\right)\right]\nonumber\\
%	\langle\sigma v_{rel}\rangle_{h\phi} &=& \frac{\beta_f^2G^2+\tilde{G}^2}{64\pi m_f^2}\sqrt{(4m_f^2+m_h^2-m_\phi^2)^2-16m_f^2m_h^2}\nonumber\\
%	& &\left[\left(\frac{s_\theta\lambda_{h^2\phi}}{4m_f^2-m_h^2} + \frac{c_\theta\lambda_{h\phi^2}}{4m_f^2-m_\phi^2}\right)^2\right.\\
%	& & + \frac{64m_f^2 G^2 s_\theta^2 c_\theta^2}{(4m_f^2-m_h^2-m_\phi^2)^2}\nonumber\\
%	& &\left. - \frac{16m_f G s_\theta c_\theta}{4m_f^2-m_h^2-m_\phi^2}\left(\frac{s_\theta\lambda_{h^2\phi}}{4m_f^2-m_h^2} + \frac{c_\theta\lambda_{h\phi^2}}{4m_f^2-m_\phi^2}\right)\right]\nonumber
%\eeqa
\begin{align}
	\langle\sigma v\rangle_{\psi_i\psi_j} &= \frac{\beta_{\chi}^{2}G^{2}+\tilde{G}^{2}}{16\pi} \left(\frac{s_{\theta}y_{hij}}{s-M_{h}^{2}}+\frac{c_{\theta} y_{\phi ij}}{s-M_{\phi}^{2}}\right)^2\nonumber\\
	&\quad \times \sqrt{s-(m_{i}+m_{j})^{2}}\sqrt{\lambda\left[\frac{m_i^2}{s},\frac{m_j^2}{s}\right]},
\end{align}
where 
\begin{align}
	y_{hii} &= \frac{y}{\sqrt{2}} c_{\theta} R_{1i}R_{2i} + gs_{\theta} R_{2i}R_{3i},\\
	y_{\phi ii} &= \frac{-y}{\sqrt{2}} s_{\theta} R_{1i}R_{2i} + gc_{\theta} R_{2i}R_{3i},\\
	y_{hij} &= -\frac{1}{\sqrt{2}}y c_{\theta} \left(R_{1i}R_{2j} + R_{1j}R_{2i}\right)\nonumber\\
	& \quad -g s_{\theta}\left(R_{2i}R_{3j} + R_{2j}R_{3i}\right),\\
	y_{\phi ij} &= +\frac{1}{\sqrt{2}}y s_{\theta} \left(R_{1i}R_{2j} + R_{1j}R_{2i}\right)\nonumber\\
	& \quad -g c_{\theta}\left(R_{2i}R_{3j} + R_{2j}R_{3i}\right).
\end{align}

\section{Decay width}\label{ch:decaywidth}
We discuss the decay width of the scalar mediator in this section. Due to mixing in the scalar sector, the decay width of $\phi$ into the SM particles takes the following form
\beqa
\Gamma(\phi \rightarrow VV) &=& \frac{\sin^2\theta}{32\pi}\frac{m_{\phi}^3}{v^2}\delta_V\nonumber\\
& & \times\sqrt{1-4x}(1-4x+12x^2),\\
\Gamma(\phi \rightarrow \overline{f}f) &=& \frac{N_c \sin^2\theta}{8\pi}\frac{m_{\phi}m_f^2}{v^2}\beta_f^3,
\eeqa
where $V=W,Z$, $\delta_W=2$, $\delta_Z=1$, $x = m_V^2/m^2_{\phi}$ and $\beta_f = \sqrt{1-\frac{4m_f^2}{m_{\phi}^2}}$. The decay width to neutrinos can be obtained straightforwardly
%\beqa
%\Gamma(\phi \rightarrow \nu \psi_i) &=& \frac{1}{32\pi}\lambda^2_{\phi\nu \psi_i}\sqrt{m_{\phi}^2-(m_{\psi_i}-m_{\nu})^2}\nonumber \\
%& & \times \left( 1 - \frac{(m_{\nu}+m_{\psi_i})^2}{m_{\phi}^2}\right)^{3/2},\\
%2\Gamma(\phi \rightarrow \psi_i \psi_i) &=& \Gamma(\phi \rightarrow \psi_2 \psi_3) \nonumber \\
%&=& \frac{1}{32\pi}\lambda^2_{\phi \psi \psi}m_{\phi} \left( 1 - \frac{4 m_{\psi}^2}{m_{\phi}^2}\right)^{3/2},
%\eeqa
\begin{align}
	\Gamma(\phi \rightarrow \psi_i\psi_j) &= \frac{1}{32\pi}y^2_{\phi ij}\sqrt{m_{\phi}^2-(m_{i}-m_{j})^2} \nonumber \\
	&\qquad \times \lambda\left[\frac{m_i^2}{s},\frac{m_j^2}{s}\right]^{3/2}.
\end{align}
Finally, for a sufficiently heavy $\phi$, it can decay into 2 Higgs bosons 
\begin{equation}
\Gamma(\phi\rightarrow hh) = \frac{1}{32\pi}\frac{\lambda_{\phi h h}^2}{m_{\phi}}\left(1-\frac{4m_h^2}{m_{\phi}^2}\right)^{1/2}.
\end{equation}

%%%%%%%%%%%%%%%%%%%%%%%%%%%%%%%%%%%%%%%%%%%%%%%

\bibliographystyle{apsrev4-1}

\bibliography{ref}

\end{document}